\documentstyle[11pt,aaspp4]{article}

\begin{document}

\singlespace

\title{
An Investigation of Neutrino-Driven Convection
and the Core Collapse Supernova Mechanism 
Using Multigroup Neutrino Transport
}

\author{A. Mezzacappa\altaffilmark{1,2}, 
A. C. Calder\altaffilmark{1,3}, 
S. W. Bruenn\altaffilmark{4}, 
J. M. Blondin\altaffilmark{5}, 
M. W. Guidry\altaffilmark{1,2},\linebreak
M. R. Strayer\altaffilmark{1,2}, and 
A. S. Umar\altaffilmark{3}}

\altaffiltext{1}{Theoretical and Computational Physics Group, Oak Ridge National
Laboratory, Oak Ridge, TN 37831--6354}
\altaffiltext{2}{Department of Physics and Astronomy, University of Tennessee,
Knoxville, TN 37996--1200}
\altaffiltext{3}{Department of Physics and Astronomy, Vanderbilt University,
Nashville, TN 37235}
\altaffiltext{4}{Department of Physics, Florida Atlantic University, Boca Raton, FL 33431--0991}
\altaffiltext{5}{Department of Physics, North Carolina State University, 
Raleigh, NC 27695--8202}

\begin{center}
{\large {\bf Abstract}}
\end{center}

We investigate neutrino-driven convection in core collapse supernovae and 
its ramifications for the explosion mechanism. We begin with an ``optimistic'' 
15 M$_{\odot}$ precollapse model, which is representative of the class of 
stars with compact iron cores. This model is evolved through core collapse 
and bounce in one dimension using multigroup (neutrino-energy--dependent) 
flux-limited diffusion ({\small MGFLD}) neutrino transport and Lagrangian 
hydrodynamics, providing realistic initial conditions for the postbounce 
convection and evolution. 
Our two-dimensional simulation begins at 106 ms after bounce at a time when 
there is a well-developed gain region, and proceeds for 400 ms. We couple 
two-dimensional ({\small PPM}) hydrodynamics to one-dimensional {\small MGFLD} 
neutrino transport. 

At 225 ms after bounce we see large-scale convection behind the shock, 
characterized by high-entropy, mushroom-like, expanding upflows 
and dense, low-entropy, finger-like downflows. The upflows reach the 
shock and distort it from sphericity. The radial convection velocities become
supersonic just below the shock, reaching magnitudes in excess of $10^{9}$ 
cm/sec. Eventually, however, the shock recedes to smaller radii, and at 
$\sim$500 ms after bounce there is no evidence in our simulation of an explosion or of a developing 
explosion. 

%
Failure in our ``optimistic'' 15 M$_{\odot}$ Newtonian model leads us to
conclude that it is unlikely, at least in our approximation, that neutrino-driven 
convection will lead to explosions for more massive stars with fatter iron 
cores or in cases in which general relativity is included. 

\keywords{(stars:) supernovae: general -- convection}
\newpage

\section{\bf Introduction}

Despite the best efforts of theorists over three decades, the core collapse
supernova mechanism, at least in detail, remains elusive. 
Current supernova modeling is centered around the idea that the supernova shock
wave, which stalls in the stellar core because of dissociation and neutrino losses, 
is reenergized by electron neutrino and antineutrino absorption on nucleons behind 
it (Bethe \& Wilson 1985; Wilson 1985), although no recent numerical simulations 
produce explosions unless the neutrino luminosities or the energy deposition 
efficiencies are boosted by convection (Wilson \& Mayle 1993, Herant et al. 1994, 
Burrows et al. 1995, Janka \& M\"{u}ller 1996). One potentially important mode is 
neutrino-driven convection below the shock (Herant et al. 1992), which is the 
subject of this Letter. It is driven by a negative entropy gradient established 
by neutrino heating (primarily absorption) as the shocked matter infalls. 

Two-dimensional simulations of neutrino-driven convection and core collapse supernovae 
have produced mixed results. Herant et al. (1992, 1994) find that neutrino-driven convection 
consistently yields robust explosions, whereas Miller et al. (1993), Burrows et al. (1995), 
and Janka and M\"{u}ller (1996) do not. Burrows et al. (1995) point out that success or 
failure in their algorithm depends sensitively on the choice of neutrino--matter coupling 
above the neutrinospheres; Janka and M\"{u}ller have shown systematically that neutrino-driven 
convection only aids in generating explosions for a narrow range of neutrino luminosities; 
Miller et al. (1993) find that neutrino-driven convection develops too slowly to be relevant 
for the postbounce supernova evolution. 

These disparate outcomes most likely result from differences in the numerical
hydrodynamics methods and neutrino transport approximations used by each group,
although differences in equations of state, neutrino opacities, etc. probably
contribute, too. Most important, 
no group has yet implemented neutrino transport that simultaneously (a) is 
multidimensional, (b) is multigroup (neutrino-energy--dependent), and (c) 
simulates with sufficient realism the transport of neutrinos in all three important regions: opaque, 
semitransparent, and transparent. 
It has been shown that the initial 
conditions for convection and the postbounce supernova evolution are sensitive to 
the sophistication of neutrino transport during core collapse and bounce,
with greater sophistication leading to weaker shocks and the establishment of 
smaller initial entropy gradients to drive convection [e.g., see Bruenn \& 
Mezzacappa (1994)]. 
Moreover, it is well established that the neutrino shock-reheating mechanism is 
extremely sensitive to the luminosities, spectra, and flux factors of the electron 
neutrinos and antineutrinos that emerge from the neutrinospheres [e.g., see Burrows 
et al. (1995) and Janka \& M\"{u}ller (1996)]. 

Our goal in this Letter is to fulfill criteria (b) and (c) in the context 
of multidimensional supernova modeling: We couple one-dimensional multigroup 
flux-limited diffusion neutrino transport to two-dimensional hydrodynamics. 
This satisfies (b) and (c), and is the first implementation of multigroup 
transport in this context. 

\section{Initial Models, Codes, and Methodology}

We begin with the 15 ${\rm M}_{\odot}$ precollapse model S15s7b 
provided by Woosley (1995). The initial model was evolved through 
core collapse and bounce using {\small MGFLD} neutrino transport 
and Lagrangian hydrodynamics, providing realistic initial conditions
for the postbounce convection and evolution. The one-dimensional data at 106 ms 
after bounce (305 ms after the initiation of core collapse) were 
mapped onto our two-dimensional Eulerian grid. We selected an
initial postbounce slice that had a well-developed gain region. 
The inner and outer boundaries of our grid were chosen to be at 
radii of 20 km and 1000 
km, respectively. 128 nonuniform radial spatial zones were used, 
and 128 uniform angular zones spanning a range of 180 degrees were 
used for $\theta$ (together with reflecting boundary conditions). 
[For more detail, see Mezzacappa et al. (1996a) and Calder 
et al. (1996ab).] 

The initial Ledoux (entropy and electron fraction) unstable region below the shock at the onset of our 
simulation extended from a radius of 99 km to a radius of 170 km. At the base of the unstable 
region, the electron fraction, $Y_{\rm e}$, is equal to 0.2375; at the top, $Y_{\rm e}
=0.4956$. The maximum entropy per baryon (in units of Boltzmann's 
constant) is 10.66, and drops to 8.574 at the top of the region.

When the matter in our simulations is in nuclear statistical equilibrium ({\small NSE}), 
we describe its thermodynamic state using the Lattimer--Swesty equation 
of state (Lattimer \& Swesty 1991). At late times in our outermost zones, 
the densities and/or temperatures become low enough that the infalling matter 
is no longer in {\small NSE}. In this instance, i.e., 
when our deflashing threshold ($\rho <1.674\times 10^{7}$ g/cm$^{3}$ and/or 
$T<0.3447$ MeV) is crossed in any given zone, the zone is deflashed to 
silicon in an energy conserving way. An ideal gas equation of state is 
then used to describe the silicon in its subsequent evolution, which
includes internal degrees of freedom to mimic the Lattimer-Swesty
equation of state to provide a seemless thermodynamic transition 
between {\small NSE} and non-{\small NSE}.

In our two-dimensional simulations, the Newtonian gravity was assumed 
to be spherically symmetric. 
The gravitational field in the convectively unstable region was dominated 
by the enclosed mass at the region's base, which at the start of our simulations 
was 1.33 M$_{\odot}$; at this time, the enclosed mass at the top of the region 
was 1.36 M$_{\odot}$. 

Details of our codes and more detail on our methodology can be found in 
Mezzacappa et al. (1996a) and Calder et al. (1996ab).

\section{\bf Results}

Figure 1 illustrates the evolution of entropy and neutrino-driven convection 
during the course of our two-dimensional 15 M$_{\odot}$ simulation. At 
$t=424$ ms (i.e., 119 ms from the start of our simulation and 225 
ms after bounce), $R_{\rm shock}\approx 205$ km, and neutrino-driven convection 
is fully developed. Semiturbulent, large-scale, convective flows below the shock are evident, 
qualitatively similar to the convection seen by Burrows et al. (1995) and 
Janka and M\"{u}ller (1996) using the same numerical hydrodynamics method ({\small PPM}), 
but more turbulent than the convective flows seen by Herant et al (1994) using 
smooth particle hydrodynamics. High-entropy, mushroom-like, rising plumes 
reach the shock and distort it from sphericity, while material behind the shock 
infalls in dense, low-entropy, fingers. 

At $t=424$ ms, the angle-averaged entropy defined by $\langle s\rangle (r)={\sum_{\theta}s(r,\theta )
A(r,\theta )}/A$, where $A=4\pi r^2$ and $A(r,\theta )=2\pi r^{2}\sin\theta\Delta\theta$, 
has a maximum $s_{\rm max}=13.5$. The convection is subsonic over most of the convecting 
region but becomes supersonic just below the shock. The average radial convection velocities, 
defined by $\langle v^{\rm C}_{r}\rangle (r) ={\sum_{\theta}[||v_{r}(r,\theta )|- \langle 
v_{r}\rangle (r)|]A(r,\theta )}/A$, where $\langle v_{r}\rangle (r)={\sum_{\theta}|v_{r}
(r,\theta )|A(r,\theta )}/A$, reach magnitudes in excess of $10^{9}$ cm/sec [for 
details, see Calder et al. (1996b)]. 

Despite the development of significant neutrino-driven convection below the shock, the shock 
eventually recedes to smaller radii from $\sim 200$ km at $t=204$ ms after bounce to $\sim 100$ km at 
$t=506$ ms after bounce, 
and the convection becomes more turbulent. Our simulation ends at $t=705$ ms (506 ms after 
bounce), with no evidence of an explosion or of a developing explosion. Other groups obtain 
explosions within the first 50--100 ms after bounce (Herant et al. 1994; Burrows et al. 1995; 
Janka \& M\"{u}ller 1996). 

The neutrino heating rate (in MeV/nucleon) in the region between the neutrinospheres 
and the shock can be written as

\begin{equation}
\dot{\epsilon}=\frac{X_{n}}{\lambda_{0}^{a}}\frac{L_{\nu_{\rm e}}}{4\pi r^{2}}
                \langle\epsilon^{2}_{\nu_{\rm e}}\rangle\langle\frac{1}{f}\rangle
              +\frac{X_{p}}{\bar{\lambda}_{0}^{a}}\frac{L_{\bar{\nu}_{\rm e}}}{4\pi r^{2}}
                \langle\epsilon^{2}_{\bar{\nu}_{\rm e}}\rangle\langle\frac{1}{\bar{f}}\rangle
\end{equation}
\smallskip

\noindent where $X_{n,p}$ are the neutron and proton fractions; $\lambda_{0}^{a},
\bar{\lambda}_{0}^{a}$ are the coefficients of the $\epsilon_{\nu_{\rm e},\bar{\nu}_{\rm e}}^{-2}$ 
neutrino energy dependences in the electron neutrino and antineutrino mean free paths, respectively; 
$L_{\nu_{\rm e},\bar{\nu}_{\rm e}}$, $\langle\epsilon^{2}_{\nu_{\rm e},\bar{\nu}_{\rm e}}\rangle$, 
and $\langle 1/f,\bar{f}\rangle$ are the electron neutrino and antineutrino 
luminosities, mean square energies, and mean inverse flux factors, respectively, as defined in 
Mezzacappa et al. (1996b). Success in generating explosions by neutrino heating must ultimately 
rest on these three key neutrino quantities. 
In Figures 2 and 3, we plot them as a function of radius for select times during 
our simulation. We supply this complete set of neutrino data to facilitate comparison 
with other groups; in the past, only partial data have been made available in the 
literature. 

Equation (1) is appropriate for neutrino emission and absorption. In our simulation,
the heating contributions from neutrino--electron scattering ({\small NES}) are negligible, 
amounting to 3--5\% corrections for our postshock entropies ($\leq 17-18$). At 
typical postshock densities between $10^{8}$ and $10^{9}$ g/cm$^{3}$, entropies 
$\sim 30$ (almost twice as large as our entropies) would be required before 
the number density of pairs would become comparable to the baryon number density, 
i.e., before our {\small NES} heating contributions would double. 

\section{\bf Discussion}

Our results depend in part on the assumption that our electron neutrino 
and antineutrino sources remain to a good approximation spherically 
symmetric during the course of our two-dimensional run. This requires 
that there be no significant convection in the region encompassing or 
below the neutrinospheres and no significant influence of neutrino-driven
convection below the gain radii:

(A) {\it Convection Below the Neutrinospheres:}~ 
In a previous paper (Mezzacappa et al. 1996a), we demonstrated that, in the 
presence of neutrino transport, the convective transport of heat and
leptons below the neutrinospheres by {\it prompt} convection is insignificant.
Our numerical results were supported by timescale analyses and by a simple
analytical model. These results are mentioned here in support of the 
conclusions reached in this Letter. In the absence of prompt convection,
the imposition of a one-dimensional spherically-symmetric neutrino
radiation field in the region between the neutrinospheres and the
shock, which is used to compute the neutrino heating and cooling there, 
is a better approximation. 

(B) {\it The Influence of Neutrino-Driven Convection Below the Gain Radii:}~ 
Because our current prescription does not implement a self-consistent two-dimensional 
radiation hydrodynamics solution, we cannot capture enhancements in the neutrino 
luminosities emanating from the neutrinosphere region that result from (1) non--spherically-symmetric
accretion through the gain radius and/or (2) inwardly propagating nonlinear 
waves that compress and heat the neutrinosphere region in a non--spherically-symmetric 
way. For example, the dense, finger-like, low-entropy inflows in the neutrino-driven
convection region may penetrate the gain radius and strike the protoneutron star 
surface (Burrows et al. 1995, Janka \& M\"{u}ller 1996). It has been suggested 
that the associated luminosity enhancements 
may help trigger explosions (Burrows et al. 1995), but conclusions regarding 
their benefit have been mixed (Janka \& M\"{u}ller 1996). 

To investigate whether these effects would have been important in 
our simulation, we compared our one- and two-dimensional density, temperature, and 
electron fraction snapshots at $t=424$ ms, a time when neutrino-driven convection 
was most vigorous. Up to the neutrinosphere radii ($\sim 50$ km), we 
found no differences. Between the neutrinospheres
and the gain radii ($\sim 85$ km), we found hot spots in our two-dimensional
simulation where $\Delta T/T\sim 3\%$ over $\sim 1/4-1/3$ of the 
volume, and $Y_{\rm e}$-enhanced spots where $\Delta Y_{\rm e}/Y_{\rm e}
\sim 3-6\%$ over $\sim 1/5$ of the volume. Because of the $T^{6}$ dependence
in the neutrino emission rates, to first order we would expect localized luminosity
enhancements $\sim 18\%$. However, at $t\sim 100-200$ ms after bounce, $L_{\nu_{\rm e}}(50\, 
{\rm km})\approx 2.4\times 10^{52}$ erg/sec and  $L_{\nu_{\rm e}}(90\, {\rm km})\approx 
3.5\times 10^{52}$ erg/sec; therefore, only $\sim 33\%$ of the neutrino luminosities would have
been affected by these temperature and electron-fraction enhancements. (The percentages 
for electron antineutrinos are comparable: $\sim 50\%$ at 100 ms and $\sim 33\%$ at 200
ms.)

%
%
%
Considering the small local enhancements in $T$ and $Y_{\rm e}$, the small
percentage of the volume in which they occur, and the fraction of
the neutrino luminosities that would be affected by them, we do not expect 
these enhancements to have significant ramifications for the
supernova outcome.

\section{\bf Summary and Conclusions}

With two-dimensional ({\small PPM}) hydrodynamics coupled to one-dimensional
multigroup flux-limited diffusion neutrino transport, we see vigorous --- in some
regions supersonic --- neutrino-driven convection develop behind the shock, but 
despite this, do not obtain explosions for what should be an ``optimistic'' 
15 M$_{\odot}$ model. Beginning with realistic postbounce initial conditions, 
our simulation has been carried out for $\sim$500 ms, a period that is long 
relative to the 50--100 ms explosion timescales obtained by other groups.

%
We have considered the non--spherically-symmetric luminosity enhancements
that would occur from local temperature and electron fraction enhancements 
below the gain radii (which enclose the electron neutrino and antineutrino 
sources) that are seen in our two-dimensional run, which result either from non--spherically-symmetric
accretion through the gain radius or nonlinear inwardly propagating 
non--spherically-symmetric waves. We see no enhancements below the 
neutrinosphere radii; between them and the gain radii, we see small 
enhancements that occur over a small fraction of the volume responsible 
for producing less than one third of the neutrino luminosities. From this, 
we conclude that the use of one-dimensional {\small MGFLD} neutrino transport in our two-dimensional simulations 
is a good approximation.

We do not expect to obtain explosions for more massive stars. [The results 
for other models will be reported in Calder et al. (1996b).] Moreover, our 
simulations are Newtonian; with general relativistic gravity, conditions 
will be even more pessimistic. The neutrino luminosities will be redshifted,
the increased infall velocities and the smaller width between the gain radii 
and the shock will allow less time for neutrino heating to reverse infall, 
and everything will occur in a deeper gravitational well, making explosion
more difficult.
  
Our results point to the need, at least in our approximation, for either improved neutrino 
transport (relative to {\small MGFLD}) or new physics in order to obtain consistent 
robust explosions. Recently, we have obtained new results from comparisons
of three-flavor Boltzmann neutrino transport and three-flavor {\small MGFLD} 
in postbounce supernova environments (thermally frozen, hydrostatic). In 
particular, the Boltzmann electron neutrino and antineutrino heating rates 
between the neutrinospheres and the shock are larger, and the Boltzmann net 
heating rates in the region directly above the gain radii are significantly 
larger (Mezzacappa et al. 1996b). These results suggest that Boltzmann transport 
will yield greater neutrino heating and more vigorous neutrino-driven convection; 
both would increase the chances of reviving the stalled shock. 

\section{Acknowledgements}

AM, ACC, MWG, and MRS were supported at the Oak Ridge National 
Laboratory, which is managed by Lockheed Martin Energy Research 
Corporation under DOE contract DE-AC05-96OR22464. AM, MWG, and 
MRS were supported at the University of Tennessee under DOE contract 
DE-FG05-93ER40770. ACC and SU were supported at Vanderbilt 
University under DOE contract DE-FG302-96ER40975. SWB was 
supported at Florida Atlantic University under NSF grant 
AST--941574, and JMB was supported at North Carolina State 
University under NASA grant NAG5-2844. 
The simulations presented in this Letter were carried out on
the Cray C90 at the National Energy Research Supercomputer
Center, the Cray Y/MP at the North Carolina Supercomputer 
Center, and the Cray Y/MP and Silicon Graphics Power
Challenge at the Florida Supercomputer Center. We would 
like to thank 
Willy Benz, 
Stirling Colgate, 
Thomas Janka, 
Michael Smith, 
Jim Stone, 
and 
Friedel Thielemann 
for stimulating discussions. 
\newpage

\section{\bf References}

\begin{enumerate}
\item Bethe, H. \& Wilson, J. R. 1985, ApJ, 295, 14
\item Bruenn, S. W. \& Mezzacappa, A. 1994, ApJ, 433, L45
\item Burrows, A., Hayes, J., \& Fryxell, B. A. 1995, ApJ, 450, 830
\item Calder, A. C., Mezzacappa, A., Bruenn, S. W., Blondin, J. M., 
      Guidry, M. W., Strayer, M. R., \& Umar, A. S. 1996a, ApJ, in preparation
\item Calder, A. C., Mezzacappa, A., Bruenn, S. W., Blondin, J. M., 
      Guidry, M. W., Strayer, M. R., \& Umar, A. S. 1996b, ApJ, in preparation
\item Herant, M. E., Benz, W., Hix, W. R., Fryer, C., \& Colgate, S. A. 1994, 
      ApJ, 435, 339
\item Herant, M. E., Benz, W., \& Colgate, S. A. 1992, ApJ, 395, 642
\item Janka, H.-Th., \& M\"{u}ller, E. 1996, A\& A 306, 167 
\item Lattimer, J. M. \& Swesty , F. D. 1991, Nucl. Phys. A, 535, 331
\item Mezzacappa, A., Messer, O. B., Bruenn, S. W., \& Guidry, M. W. 1996b, 
      ApJ, in preparation 
\item Mezzacappa, A., Calder, A. C., Bruenn, S. W., Blondin, J. M., 
      Guidry, M. W., Strayer, M. R., \& Umar, A. S. 1996a, ApJ, submitted 
\item Miller, D. S., Wilson, J. R., \& Mayle, R. W. 1993, ApJ, 415, 278
\item Wilson, J. R. \& Mayle, R. W. 1993, Phys. Rep., 227, 97 
\item Wilson, J. R. 1985, in Numerical Astrophysics, eds. J. M. Centrella et al.
      (Boston: Jones \& Bartlett), 422 
\item Woosley, S. E. 1995, private communication
\end{enumerate}
\newpage

\figcaption{The two-dimensional entropy at three ``early'' time slices for our 
15 M$_{\odot}$ model.
Large-scale convection with high-entropy expanding upflows and low-entropy dense 
downflows is evident. The upflows extend to the shock, do work, and distort it.}

\figcaption{The electron neutrino luminosity, RMS energy, and mean inverse flux
factor are plotted as a function of radius at select times during our 15 
M$_{\odot}$ simulation.}

\figcaption{The electron antineutrino luminosity, RMS energy, and mean inverse flux 
factor are plotted as a function of radius at select times during our 15 
M$_{\odot}$ simulation.}

\end{document}